\documentclass[manuscript]{aastex}

\usepackage{lineno} 

\newcommand{\blaz}{1ES~1218+304}
\newcommand{\gray}{$\gamma$-ray}
\newcommand{\grays}{$\gamma$-rays}
\newcommand{\cerenkov}{Cherenkov}
\slugcomment{To appear in \apj Letter }

\shorttitle{VARIABILITY IN THE VHE \gray{}-RAY EMISSION OF 1ES 1218$+$304 WITH VERITAS}
\shortauthors{Acciari et al.}

\begin{document}

\title{Discovery of Variability in the Very High Energy $\gamma$-Ray \\
       Emission of 1ES 1218+304 with VERITAS}

\author{
V.~A.~Acciari\altaffilmark{1},
E.~Aliu\altaffilmark{2},
M.~Beilicke\altaffilmark{3},
W.~Benbow\altaffilmark{1},
D.~Boltuch\altaffilmark{4},
M.~B{\"o}ttcher\altaffilmark{5},     %%
S.~M.~Bradbury\altaffilmark{6}, 
V.~Bugaev\altaffilmark{3},
K.~Byrum\altaffilmark{7},
A.~Cesarini\altaffilmark{8},
L.~Ciupik\altaffilmark{9},
P.~Cogan\altaffilmark{10},
W.~Cui\altaffilmark{11},
R.~Dickherber\altaffilmark{3},
C.~Duke\altaffilmark{12},
A.~Falcone\altaffilmark{13},
J.~P.~Finley\altaffilmark{11},
G.~Finnegan\altaffilmark{14},
L.~Fortson\altaffilmark{9},
A.~Furniss\altaffilmark{15},
N.~Galante\altaffilmark{1},
D.~Gall\altaffilmark{11},
K.~Gibbs\altaffilmark{1},
R.~Guenette\altaffilmark{10},
G.~H.~Gillanders\altaffilmark{8},
S.~Godambe\altaffilmark{14},
J.~Grube\altaffilmark{16},
D.~Hanna\altaffilmark{10},
C.~M.~Hui\altaffilmark{14},
T.~B.~Humensky\altaffilmark{17},
A.~Imran\altaffilmark{18,*},
P.~Kaaret\altaffilmark{19},
N.~Karlsson\altaffilmark{9},
M.~Kertzman\altaffilmark{20},
D.~Kieda\altaffilmark{14},
H.~Krawczynski\altaffilmark{3},
F.~Krennrich\altaffilmark{18},
M.~J.~Lang\altaffilmark{8},
S.~LeBohec\altaffilmark{14},
G.~Maier\altaffilmark{10},
S.~McArthur\altaffilmark{3},
A.~McCann\altaffilmark{10},
P.~Moriarty\altaffilmark{21},
T.~Nagai\altaffilmark{18},
R.~A.~Ong\altaffilmark{22},
A.~N.~Otte\altaffilmark{15},
D.~Pandel\altaffilmark{16},
J.~S.~Perkins\altaffilmark{1},
A.~Pichel\altaffilmark{23},
M.~Pohl\altaffilmark{18},
J.~Quinn\altaffilmark{16},
K.~Ragan\altaffilmark{10},
L.~C.~Reyes\altaffilmark{24},
P.~T.~Reynolds\altaffilmark{25},
E.~Roache\altaffilmark{1},
H.~J.~Rose\altaffilmark{6},
M.~Schroedter\altaffilmark{18},
G.~H.~Sembroski\altaffilmark{11},
A.~W.~Smith\altaffilmark{7},
D.~Steele\altaffilmark{9},
S.~P.~Swordy\altaffilmark{17},
M.~Theiling\altaffilmark{1},
S.~Thibadeau\altaffilmark{3},
V.~V.~Vassiliev\altaffilmark{22},
S.~Vincent\altaffilmark{14},
S.~P.~Wakely\altaffilmark{17},
T.~C.~Weekes\altaffilmark{1},
A.~Weinstein\altaffilmark{22},
T.~Weisgarber\altaffilmark{17},
D.~A.~Williams\altaffilmark{15}
}

\altaffiltext{1}{Fred Lawrence Whipple Observatory, Harvard-Smithsonian Center for Astrophysics, Amado, AZ 85645, USA}
\altaffiltext{2}{Department of Physics and Astronomy, Barnard College, Columbia University, NY 10027, USA}
\altaffiltext{3}{Department of Physics, Washington University, St. Louis, MO 63130, USA}
\altaffiltext{4}{Department of Physics and Astronomy and the Bartol Research Institute, University of Delaware, Newark, DE 19716, USA}
\altaffiltext{5}{Astrophysical Institute, Department of Physics and Astronomy, Ohio University, Athens, OH 45701, USA}
\altaffiltext{6}{School of Physics and Astronomy, University of Leeds, Leeds, LS2 9JT, UK}  
\altaffiltext{7}{Argonne National Laboratory, 9700 S. Cass Avenue, Argonne, IL 60439, USA}
\altaffiltext{8}{School of Physics, National University of Ireland Galway, University Road, Galway, Ireland}
\altaffiltext{9}{Astronomy Department, Adler Planetarium and Astronomy Museum, Chicago, IL 60605, USA}
\altaffiltext{10}{Physics Department, McGill University, Montreal, QC H3A 2T8, Canada}
\altaffiltext{11}{Department of Physics, Purdue University, West Lafayette, IN 47907, USA }
\altaffiltext{12}{Department of Physics, Grinnell College, Grinnell, IA 50112-1690, USA}
\altaffiltext{13}{Department of Astronomy and Astrophysics, 525 Davey Lab, Pennsylvania State University, University Park, PA 16802, USA}
\altaffiltext{14}{Department of Physics and Astronomy, University of Utah, Salt Lake City, UT 84112, USA}
\altaffiltext{15}{Santa Cruz Institute for Particle Physics and Department of Physics, University of California, Santa Cruz, CA 95064, USA}
\altaffiltext{16}{School of Physics, University College Dublin, Belfield, Dublin 4, Ireland}
\altaffiltext{17}{Enrico Fermi Institute, University of Chicago, Chicago, IL 60637, USA}
\altaffiltext{18}{Department of Physics and Astronomy, Iowa State University, Ames, IA 50011, USA}
\altaffiltext{19}{Department of Physics and Astronomy, University of Iowa, Van Allen Hall, Iowa City, IA 52242, USA}
\altaffiltext{20}{Department of Physics and Astronomy, DePauw University, Greencastle, IN 46135-0037, USA}
\altaffiltext{21}{Department of Life and Physical Sciences, Galway-Mayo Institute of Technology, Dublin Road, Galway, Ireland}
\altaffiltext{22}{Department of Physics and Astronomy, University of California, Los Angeles, CA 90095, USA}
\altaffiltext{23}{Instituto de Astronomia y Fisica del Espacio, Casilla de Correo 67 - Sucursal 28, (C1428ZAA) Ciudad Autónoma de Buenos Aires, Argentina}
\altaffiltext{24}{Kavli Institute for Cosmological Physics, University of Chicago, Chicago, IL 60637, USA}
\altaffiltext{25}{Department of Applied Physics and Instrumentation, Cork Institute of Technology, Bishopstown, Cork, Ireland}
\altaffiltext{*}{Corresponding author: imranisu@iastate.edu}

\begin{abstract}
We present results from an intensive VERITAS monitoring campaign of the high-frequency peaked BL Lac object \blaz{} in 2008/2009.  Although \blaz{} was detected previously by MAGIC and VERITAS at a  persistent level of $\sim$6\% of the Crab Nebula flux, the new VERITAS data reveal a prominent flare reaching $\sim$20\% of the Crab. While very high energy (VHE) flares are quite common in many nearby blazars, the case of 1ES~1218+304 (redshift $z=0.182$) is particularly interesting since it belongs to a group of blazars that exhibit unusually hard VHE spectra considering their redshifts. When correcting the measured spectra for absorption by the extragalactic background light, \blaz{} and a  number of other blazars are found to have differential photon indices $\rm \Gamma \le$~1.5. The difficulty in modeling these hard spectral energy distributions in blazar jets has led to a range of theoretical \gray{} emission scenarios, one of which is strongly constrained by these new VERITAS observations. We consider the implications of the observed light curve of \blaz{}, which shows day-scale flux variations, for shock acceleration scenarios in relativistic jets, and in particular for the viability of kiloparsec-scale jet emission scenarios.

\end{abstract}

%% Line numbering on
%%\linenumbers

\keywords{BL Lacertae objects: individual (1ES~1218+304 = VER J1221+301) ---
galaxies: active --- gamma rays: galaxies}

\section{INTRODUCTION}

The very high energy (VHE) \gray{} sky has seen a dramatic increase in the number of sources, with 26 detected active galactic nuclei\footnote{For updates see http://tevcat.uchicago.edu} (AGNs) over redshifts of 0.0018 -0.536.  All, except two, are blazars with their jets closely aligned to our line of sight. Blazars as a class are comprised of BL Lacertae objects, optically violent variables and other radio-loud quasars \citep{urr95}. Non-thermal emission dominates the broad-band continuum spectra of blazars.

 The characteristic double-peaked spectral energy distribution of blazars is usually attributed to synchrotron radiation in the radio to X-ray wavebands, and inverse-Compton up-scattering of soft target photons by relativistic electrons at GeV and TeV energies \citep{mar92,blo96}.  Alternative emission scenarios involve hadronic interactions producing neutral pions which decay into photons \citep{man92} and proton synchrotron radiation \citep{aha00}.  The latter models provide a link between AGNs and possible sources of ultra-high-energy cosmic rays.  A better understanding of the underlying acceleration and emission processes requires a larger sample of sources, which is now being provided by the {\it Fermi} Gamma-ray Space Telescope ({\it Fermi}) at 0.1~GeV to hundreds of GeV and by the imaging atmospheric \cerenkov{} telescopes (IACT), namely the MAGIC, H.E.S.S. and VERITAS Observatories, that provide coverage primarily in the $\rm 100~GeV - 10~TeV$ regime.

To date, blazars detected at VHE energies are predominantly high-frequency peaked BL Lacertae objects (HBLs).  Nevertheless, a few objects with lower peak energies, namely low-frequency-peaked BL Lacertae (LBLs; \citep{alb07,tes08}) and intermediate-frequency-peaked BL Lacertae objects (IBLs) \citep[see][]{acc08a.wc,acc08b.3c66a,acc09.wcom,ong09.1424} were detected recently by IACTs at TeV energies. The \gray{} luminosity of LBLs and IBLs generally peaks at sub-GeV to tens of GeV in energy, and they are therefore easily detectable by \textit{Fermi} while their detectability in the TeV regime by IACTs is more difficult due to their soft VHE spectra.

Among the HBLs, which are typically the domain of the TeV telescopes,  a number of objects exhibit unusually hard intrinsic power law energy spectra ($\rm dN/dE \sim E^{-\Gamma_{i}}$) after correcting for the $\rm \gamma \: \gamma \:  \rightarrow \: e^+ \: e^-$ absorption by  the cosmological diffuse EBL radiation field. While the measured spectral indices, $\rm \Gamma_m$, of these blazars (1ES~1101--232, 1ES~0347--121, 1ES~0229+200, 1ES~1218+304, RGB~J0710+591) range from 2.5 to 3.1 \citep[see][]{aha06.1101,aha07,aha07.0347,alb06.1218,acc09.1218,acc09.perkins}, the absorption-corrected spectral indices suggest very hard intrinsic source spectra in the VHE regime with $\rm \Gamma_i  \le 1.28 \pm 0.28$  \citep{kre08}, based on recent lower limits to the EBL from galaxy counts \citep{lev08}.

Similar hard  energy spectra are found by \textit{Fermi}, where, for example, the measured spectral indices for RGB~J0710+591 and 1ES~1440+122 are found to be $\rm \Gamma = 1.21 \pm 0.25$ and $\rm 1.18 \pm 0.27$ over energies from 0.2 GeV to tens of GeV \citep{abd09.fegan}.  While the findings of very hard VHE spectra are based on EBL constraints from galaxy counts, the \textit{Fermi} spectra directly resemble the intrinsic source spectra since absorption by the EBL is minimal for the covered energy range. The photon spectrum of 1ES~1218+304 measured by \textit{Fermi} yields a spectral index of $\rm \Gamma = 1.63 \pm 0.12$ between  0.2~GeV and $\rm \sim$300~GeV, but absorption may already play a role at the high-energy end \citep{abd09.fegan}.

Diffusive shock-acceleration theory \citep[for a review see][]{mal01} generally yields a limit to the spectral index of GeV -- TeV photon spectra resulting from inverse Compton scattering of $\rm \Gamma_i \ge 1.5$. Only recently,   numerical studies by Stecker et al. (2007) indicate that sufficiently hard electron spectra could be generated by diffusive shock acceleration in relativistic shocks, for the production of photon spectra  $\rm  1.0 < \Gamma_i < 1.5$. However,  \citet{bot08} suggest that even for a hard-spectrum electron population, in the framework of  a synchrotron self-Compton (SSC) scenario,  the resulting GeV--TeV \gray{} spectra should experience substantial softening from  Klein--Nishina effects making \gray{} spectra with $\rm \Gamma_i < 1.5$ less likely.

Other approaches to explain the hard \gray{} spectra are offered by ad hoc assumptions about the electron distribution, an additional absorption component in the source, or a postulate of new physics describing the propagation of \grays{} on extragalactic distance scales.  \citet{kat06} invoke a high low-energy cutoff in the electron distribution that could give the appearance of a hard \gray{} spectrum for a given energy regime.  \citet{aha8.narrowband} show that $\rm \gamma \gamma$ absorption in the source due to a narrow-band emission component from the AGN could lead to unusually hard VHE spectra.  Finally,  \citet{san09} suggested an axion like particle (ALP) that would distort the \gray{} spectrum through ALP/photon mixing on cosmological distances in the presence of intergalactic magnetic fields.

 A more easily testable model that involves known physics was recently proposed by \citet{bot08}. The authors attribute the hard photon spectra to inverse-Compton up-scattering of ambient photons from the cosmic microwave background (CMB), occurring in a kiloparsec-scale jet. In this case, a substantial fraction of the jet power is transported by hadrons to the outer regions of the jet, where it is dissipated into ultra-relativistic electrons. Inverse-Compton scattering in the Klein--Nishina regime is no longer the limitation at the highest energies since low-energy target photons from the CMB are abundant. The low magnetic field in the large-scale jet also avoids the overproduction of synchrotron radiation and allows Compton emission to dominate. Furthermore, the synchrotron emission in the radio to X-ray wavebands is assumed to originate on sub-parsec scales.  A similar mechanism was previously suggested to explain the hard X-ray spectra observed from the kiloparsec-scale jets of radio quasars \citep{tav07}.

To date, all of the blazars exhibiting very hard spectra appear to emit at a baseline level that is consistent with this picture.  However, it is also possible that the sensitivity of the current IACTs is the limiting factor in detecting day scale variations and that the underlying emission may contain flares.   The question as to whether or not the \gray{} emission from  1ES~1218+304 consists of flares or corresponds to a constant baseline emission level was one of the main motivations for VERITAS to monitor this object over a period of 5 months in the 2008/2009 season.   The other motivation for deep exposures of relatively large redshift ($z=0.1 - 0.2$) blazars at TeV energies is to provide better constraints on the EBL spectrum. The observed VHE photon spectrum can be used to constrain the EBL in the near to mid-IR. This is particularly promising for hard-spectrum blazars that provide the best sensitivity to any possible absorption feature.

\section{OBSERVATIONS AND DATA ANALYSIS }

VERITAS is an array of four 12 m IACTs located at the base camp of the F.L. Whipple Observatory in southern Arizona (31\fdg68N, 110\fdg95W, 1.3 km above sea level). Each VERITAS focal plane instrument has a 3\fdg{}5 field of view and consists of 499 photomultiplier tubes. The stereoscopic system has been fully operational since mid-2007 and is capable of detecting sources with a flux of 1\% of the Crab Nebula in $< 50$ hr of observation. In Summer 2009, Telescope 1 was relocated and the improved array configuration allows us to detect a point source at 1\% of the Crab Nebula flux under 30 hr of observations. The large effective area ($\rm \sim10^{5}~m^{2}$) of the array enables VERITAS to be sensitive over a wide range of energies (from 100 GeV to 30 TeV) with an energy resolution of $15\%-20\%$ above 300 GeV. For further details about VERITAS, see e.g., \citet{hol06}.

\blaz{} was observed by VERITAS from 2008 December to 2009 May. All data were taken in {\it wobble} mode where the source is positioned 0\fdg5 from the center of the camera in order to allow simultaneous background estimation. A quality selection was applied to the data to reject observations affected by poor weather conditions and other anomalies, resulting in a total live time of 27.2 hr for analysis. The mean zenith angle of the observations was $12\degr$.

\begin{figure}
  \centering
  \includegraphics[width=4.0in]{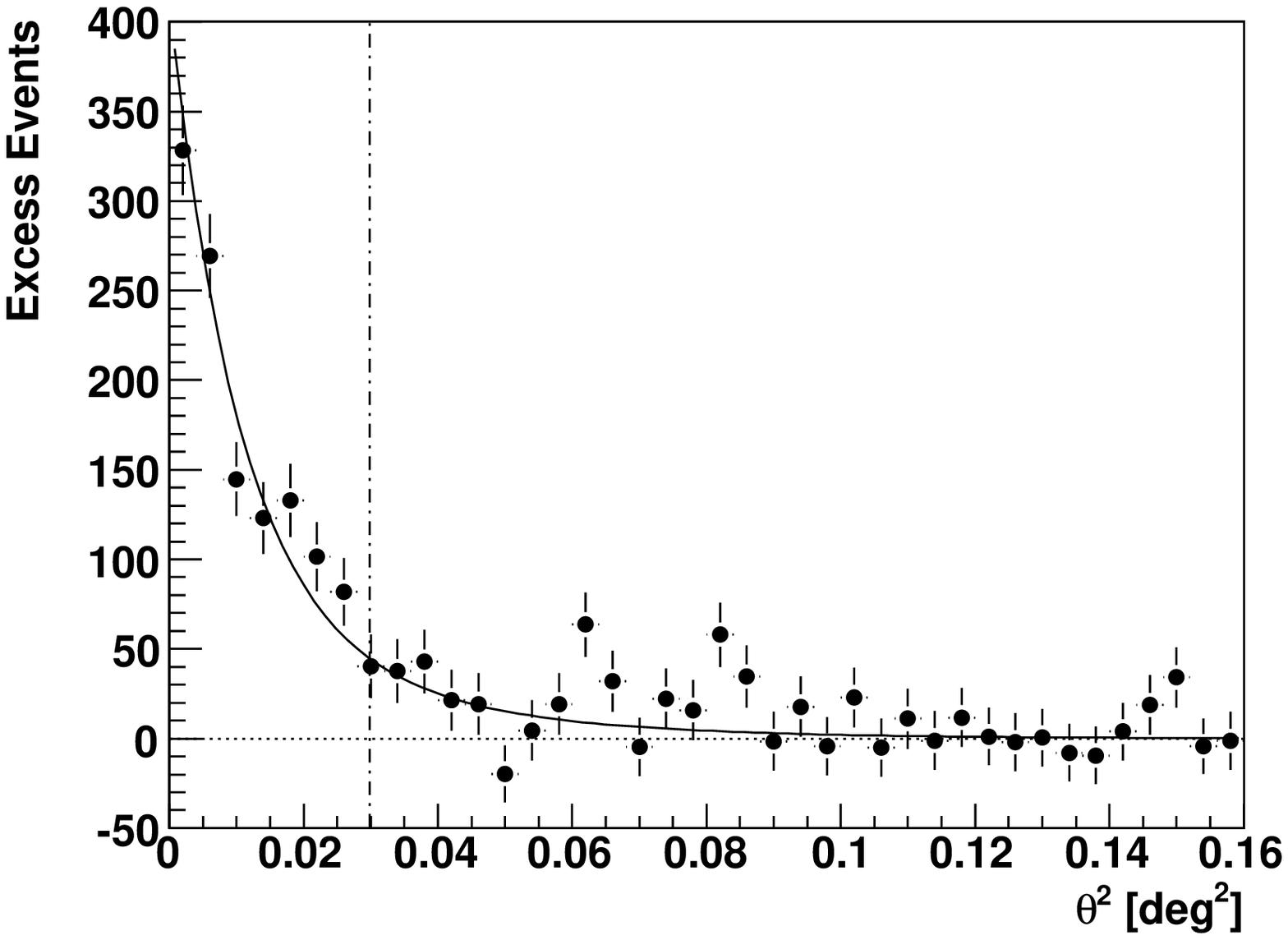}
  % \vspace{-0.1in}
  \caption{Distribution of $\theta^{2}$ for excess events ON--OFF from the observation of 1ES 1218+304. The dashed-dotted line shows the boundary of the region for the spectrum cuts ($\theta^{2}=0.03^{\circ}$). The solid curve indicates the expected  $\theta^{ 2}$ distribution from a point source and provides a good fit to the data.
    \label{t2_fig}}
\end{figure}

The data are analyzed using standard VERITAS calibration and analysis tools. Initially, the observed data set is calibrated by flat-fielding the camera gains with nightly laser runs. Following calibration and cleaning, the shower images are parameterized using a second-moment analysis \citep{hillas85}. Images from individual telescopes are combined to reconstruct both arrival direction and impact position for each event. The \gray{}-like events are separated from the cosmic ray background events  using cuts on the {\it mean-reduced-scaled length} and {\it mean-reduced-scaled width} parameters (see \citealt{aha06.crab} for a full description of the parameters). The cuts for \gray{}/hadron separation are optimized a priori for a source with a 10\% Crab-like flux. A circular region of radius $\theta$ is applied around the target to estimate the ON-source events. The background or OFF-source counts are estimated with off-source {\it reflected regions} \citep{ber07}. The significance of any excess (ON--OFF) events is calculated following Equation (17) in Li \& Ma (\citeyear{lim83}). The angular distribution of excess events from the direction of \blaz{} (Figure~\ref{t2_fig}) is comparable with the expectation from the Monte Carlo (MC) simulation of a point source.

MC simulations are used to generate a lookup table for estimating the energies of the primary \grays{} as a function of the image size and impact parameters. For spectral analysis, a looser cut ({\it spectrum cuts}) on the arrival direction of \grays{} (a region with $\theta^{2} = 0\fdg03$) is applied along with three off-source {\it reflected-regions}. This has the benefit of retaining a higher number of \gray{} candidates in addition to reducing the systematic uncertainties associated with the MC detector model. Both the standard cuts ($\theta^{2} = 0\fdg0225$) and the spectrum cuts yield compatible results. Consistent results are attained with independent analysis packages.

\section{RESULTS}

\begin{figure}
  \centering
  \includegraphics[width=5.0in]{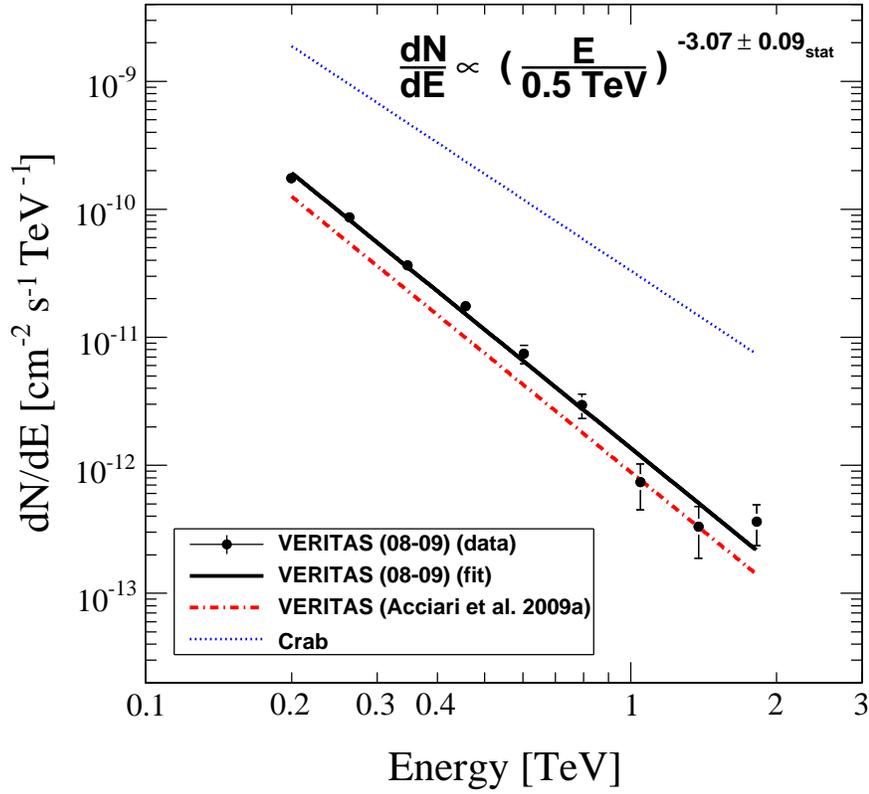}
  \caption{Observed differential energy spectrum of 1ES 1218+304 from the 2008-2009 data set. The solid line shows the power law fit to the data (filled circles) with a spectral index value of $\rm -3.07\pm0.09_{{\hspace{0.05cm}stat}}$. This agrees, within statistical uncertainties, with the previous VERITAS measurement of $\rm -3.08\pm0.34_{{\hspace{0.05cm}stat}}$ (dashed line)~\citep{acc09.1218}.
    \label{spectrum_fig}}
\end{figure}

Table~\ref{summary_tb1}  summarizes the results of the VERITAS observations of \blaz. For the spectral analysis, we report an excess of 1155 events with a statistical significance of 21.8 standard deviations, $\sigma$, from the direction of \blaz{} during the 2008-2009 campaign (2808 signal events, 4959 background events with a normalization of 0.33). Figure~\ref{spectrum_fig} shows the corresponding time-averaged differential energy spectrum. The spectrum extends from 200 GeV to 1.8 TeV and is well described ($\chi^{2}/$dof = 8.2/7) by a power law,

\begin{equation}
 \frac{dN}{dE} =  (11.5\pm0.7)\times 10^{-12}
 ~(\frac{E}{\rm{500 GeV}})^{-3.07\pm0.09_{\rm \hspace{0.05cm}stat}} {~\rm cm^{-2}s^{-1}{TeV^{-1}}}
 \label{power_eqn}
\end{equation}
The errors bars are statistical only.

Figure~\ref{lcspring_fig} shows the light curve of \blaz{} for the 2008-2009 season (binned by night). The average integral photon flux above 200 GeV is $\rm \Phi(E>200~GeV) = (18.4 \pm 0.9_{{\hspace{0.05cm}stat}})\times 10^{-12}~cm^{-2}~s^{-1}$ corresponding to $\sim$7\% of the Crab Nebula flux above the same threshold. The light curve from \blaz{} shows strong nightly flux variations between 2009 January 25 and 2009 February 5. The highest \gray{} flux is observed on MJD 54861 (2009 January 30) and is $\sim$20\% of the  Crab Nebula flux. The inset within Figure~\ref{lcspring_fig} shows the light curve above 200 GeV for nights with the highest flux. A fit of a constant to the emission during the entire season yields a $\chi^{2}/$dof of 106/21 (Figure~\ref{lcspring_fig}, dashed dotted line), corresponding to a very small $\chi^{2}$ probability ($\sim10^{-13}$). The data taken during the active period in January/February of 2009 indicate variability on a time-scale of days.

In order to investigate possible changes to the spectral shape during the flaring activity, the total  data set is divided into high-state and low-state samples. The two nights with the highest flux during flaring activity (January 30 and 31) are considered the high state, while the remaining nights are considered the low state. Both the high-state and low-state spectra are well described by a power law, and Table~\ref{states_tb2} summarizes the spectral results. We do not see any evidence for a change in the spectral index during the increase in the absolute flux level from \blaz.

\section{DISCUSSION AND CONCLUSIONS}

A prominent feature in the emission properties of blazars is the observed variability on time scales ranging from minutes to years. In 2006, {\it Suzaku} detected a prominent X-ray flare from \blaz{} revealing a hard-lag variation  in the X-rays \citep{sat08}. While \blaz{} was flagged as variable in the \textit{Fermi} bright source catalog \citep{abd09.hansen}, no variability was detected in the \textit{Fermi} light curve \citep{abd09.fegan}. The increased flux measured by VERITAS is the first significant observation of flaring activity in the VHE emission from \blaz{}. Among the other hard-spectrum VHE blazars (1ES 1101--232, 1ES 0347--121, 1ES 0229+200, 1ES1218+304, RGB J0710+591) measured to date, \blaz{} is the first clear example of significant VHE variation from a baseline flux. Consequently, the strong flare  will challenge existing spectral models of \blaz{} that assume a constant VHE emission from the source \citep[see, for example,][]{rug09}.

\begin{figure}
  \begin{center}
    \includegraphics[width=5.5in]{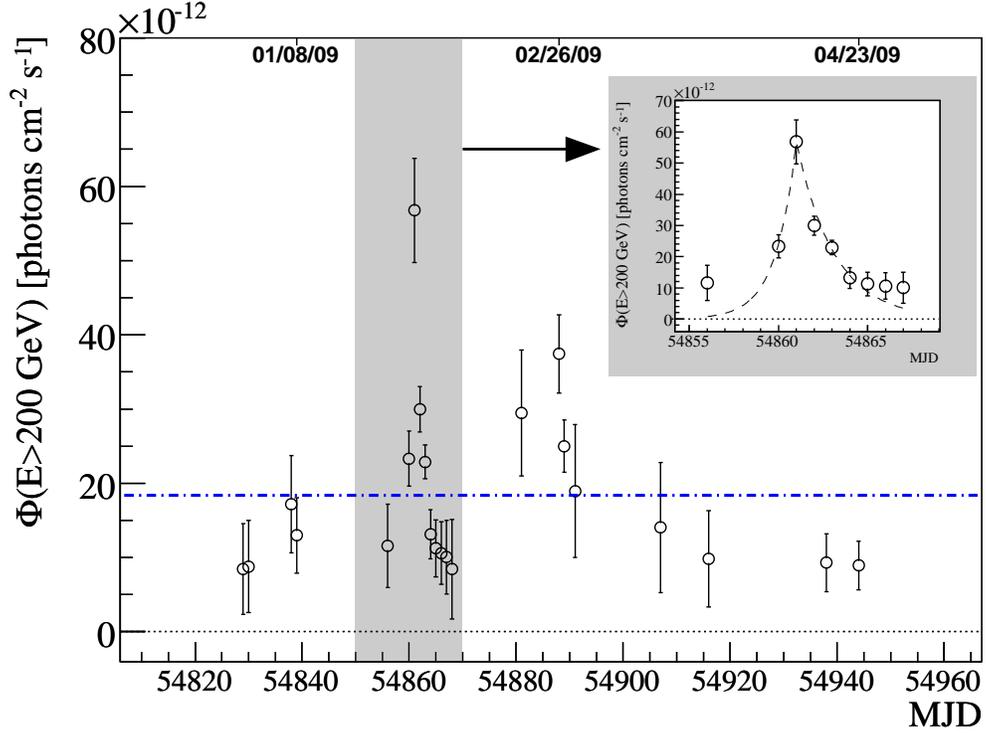}
  \end{center}
  \caption {Night-by-night VHE light curve for \blaz{} is shown as measured from 2008 December 29 (MJD 54829) to 2009 April 23 (MJD 54944). The open circles represent the integral flux above 200 GeV, $\phi({\rm{E>200~GeV}})$  from \blaz{} assuming a spectral shape, $dN/dE \propto E^{-\Gamma}$ with $\Gamma = 3.07$. The dashed-dotted line corresponds to the average integral flux of $\rm \Phi(E>200~GeV) = 18.4 \pm 0.9_{{\hspace{0.05cm}stat}}\times 10^{-12}~cm^{-2}~s^{-1}$. The inset shows the flux variations for the flaring nights in more detail. An exponential function, $e^{\lambda{}t}$ (dashed line) is used to describe the rise and fall time for flux variations giving $\lambda_{rise} = 0.86$ ($\rm\chi^{2}/dof = 3.7/2$) and $\lambda_{fall} = -0.47$ ($\rm\chi^{2}/dof = 7.2/6$), respectively. The characteristic flux doubling time, $t_{var}$ is estimated to be on the order of $\sim\ 1$ day. 
    \label{lcspring_fig}}
\end{figure}

It is generally assumed that the VHE emission from blazars is produced in the relativistic jet where the highly collimated and Doppler-boosted emission is responsible for the high luminosities and short variability timescales. The size of the emission region, $R$ is constrained by the causality argument to $R \leq ct_{var}\delta/(1+z)$ where $\delta$ is the relativistic Doppler factor and $t_{var}$ is the observed \gray{} variability timescale. An estimate of the flux-doubling time ($t_{var}\lesssim1$ day, see Figure~\ref{lcspring_fig}) limits the size of the emission region to $R\delta^{-1} \leq  2.19~\times~10^{15}~\rm cm \sim\ 0.71~\times~10^{-3}~pc$. The Doppler factor ($\delta\sim20$) typically derived for blazars \citep{mar06} implies that $R\lesssim\ 0.01~\rm pc$.

While the hard emission spectrum during the quiet state is consistent with the CMB-inverse-Compton interpretation, the requirement for a large emission region ($\rm\sim$pc~scale) predicted by the model is excluded by the $R\lesssim\ 0.01~\rm pc$ constraint imposed by the variability. Therefore, the flaring behavior from \blaz{} implies that the CMB-inverse-Compton model of emission in the kiloparsec-scale extended jet is unlikely to be the sole explanation for the extreme hardness in the intrinsic spectrum of \blaz{}.

Although our observations of \blaz{} are consistent with a baseline level of VHE emission that could have its origin in an extended emission region, the variability described here indicates that comparable flux with a comparably hard spectrum is emitted from the compact regions of this source. Further observations of \blaz{} are required to determine if the hard-spectrum emission can be explained as originating from the kiloparsec extended jet via the CMB-inverse-Compton model. Consequently, long-term VHE monitoring with ground-based \gray{} telescopes may be crucial to unraveling the emission mechanism in distant the hard-spectrum VHE blazars.

\acknowledgments
This research is supported by grants from the U.S. Department of Energy, the U.S. National Science Foundation, the Smithsonian Institution, by NSERC in Canada, by STFC in the U.K. and by Science  Foundation Ireland. We acknowledge the excellent work of the technical support staff at the FLWO and the collaborating institutions in the construction and operation of the instrument.

{\it Facilities:} \facility{VERITAS}

\bibliographystyle{apj}

\clearpage

\clearpage

% table of results
 \begin{deluxetable}{cccccc}
 \tabletypesize{\scriptsize}
 \tablecaption{Summary of observations and analysis of \blaz{}\tablenotemark{a}.\label{summary_tb1} }
 \tablewidth{0pt}
 \tablehead{
 \colhead{} &\colhead{Live Time}  & \colhead{Zenith} & \colhead{Significance} &
 \colhead{$\Phi({\rm{>200~GeV}})$} & \colhead{Units of Crab Nebula} \\
 \colhead{} & \colhead{[hours]} & \colhead{[$^{\circ}$]} & \colhead{[$\sigma$]} &
 \colhead{[$\rm 10^{-12}~cm^{-2}~s^{-1}$]} & \colhead{flux ($\rm E>200~GeV$)}
 }
 \startdata
   2006-2007\tablenotemark{b} &17.4 & 2--35 & 10.4   & $12.2\pm2.6_{{\hspace{0.05cm}stat}}$ & $0.05\pm0.011$  \\
   2008-2009       &27.2 & 2--30 & 21.8   & $18.4\pm0.9_{{\hspace{0.05cm}stat}}$ & $0.07\pm0.004$  \\

 \enddata
\tablecomments{Also shown are the integral flux above 200 GeV assuming a spectral index $\Gamma=3.07$, and the corresponding flux in units of integrated Crab Nebula flux over 200 GeV.}
\tablenotetext{a}{Including total live time, range of zenith angles for observations, and significance of the excess}
\tablenotetext{b}{\citep[see][]{acc09.1218}}
\end{deluxetable}

% table of results
 \begin{deluxetable}{cccccc}
 \tabletypesize{\scriptsize}
 \tablecaption{High-state and Low-state Data\tablenotemark{a} \label{states_tb2} }
 \tablewidth{0pt}
 \tablehead{
 \colhead{} &\colhead{Live Time}  & \colhead{$f_{0}$} & \colhead{$\Gamma$}\\
 \colhead{} &\colhead{[minute]} & &
 %\colhead{$10^{-8}~m^{-2}s^{-1}$} & \colhead{unit}
 }
 \startdata
   High     &195 & $19.4\pm2.3$  & $3.01\pm0.17$ \\
   Low      &1436 & $10.0\pm0.7$   & $3.12\pm0.10$ \\
   \tableline
   Avg      &1631& $11.5\pm0.7$  & $3.07\pm0.09$ \\

 \enddata

\tablenotetext{a}{ The high-state and low-state data (see the text for complete description) are well described by a power law fit of the form $\frac{dN}{dE} =  f_{0}\cdot(\frac{\rm{E}}{0.5~\rm{TeV}})^{-\Gamma} [\frac{10^{-12}}{\rm{cm}^{2}~\rm{s}~\rm{TeV}}]$. We note that changes to the VHE photon index during the flaring activity are statistically insignificant.}
\end{deluxetable}

\end{document}